\documentclass[a4paper,10pt,twoside]{cpc-hepnp}

\usepackage{fancyhdr}            
\usepackage{multicol}            
\usepackage{upgreek}             
\usepackage{indentfirst}         
\usepackage{booktabs}            
\usepackage{array,tabularx}      
\usepackage{keyval,graphicx}     
\usepackage{textcomp}            
\usepackage{amssymb,bm,mathrsfs,bbm,amscd} 
\usepackage[tbtags]{amsmath}     
\usepackage{lastpage}            
\usepackage[numbers,sort&compress]{natbib}  
\usepackage{color}

\begin{document}

\setlength{\abovecaptionskip}{4pt plus1pt minus1pt}   
\setlength{\belowcaptionskip}{4pt plus1pt minus1pt}   
\setlength{\abovedisplayskip}{6pt plus1pt minus1pt}   
\setlength{\belowdisplayskip}{6pt plus1pt minus1pt}   
\addtolength{\thinmuskip}{-1mu}            
\addtolength{\medmuskip}{-2mu}             
\addtolength{\thickmuskip}{-2mu}           
\setlength{\belowrulesep}{0pt}          
\setlength{\aboverulesep}{0pt}          
\setlength{\arraycolsep}{2pt}           

\providecommand{\e}[1]{\ensuremath{\times 10^{#1}}}


\fancyhead[c]{\small Chinese Physics C~~~Vol. 37, No. 6 (2013) 066002}
\fancyfoot[C]{\small 066002-\thepage}

\footnotetext[0]{Received 11 July 2012}

\title{\boldmath A normal-pressure MWPC for beam diagnostics at RIBLL2\thanks{Supported by
National Natural Science Foundation of China (11079044) }}

\author{%
TANG Shu-Wen$^{1,2;1)}$\email{tangsw\oa impcas.ac.cn}%
\quad MA Peng$^{1}$
\quad DUAN Li-Min$^{1}$\\
\quad SUN Zhi-Yu$^{1}$
\quad LU Chen-Gui$^{1}$
\quad YANG He-Run$^{1}$\\
\quad HU Rong-Jiang$^{1}$
\quad HUANG Wen-Xue$^{1}$
\quad XU Hu-Shan$^{1}$
}

\maketitle

\address{%
\vspace{1mm}
$^1$ Institute of Modern Physics, Chinese Academy of Sciences, Lanzhou 730000, China\\
$^2$ University of Chinese Academy of Sciences, Beijing 100049, China\\
}

\begin{abstract}
A normal pressure MWPC for beam diagnostics at RIBLL2 has been
developed, which has a sensitive area of 80~mm$\times$80~mm and
consists of three-layer wire planes. The anode plane is designed
with a wider frame to reduce the discharge and without using
protection wires. The detector has been tested with a $^{55}$Fe
X-ray source and a 200~MeV/u $^{12}$C beam from CSRm. A position
resolution better than 250~$\upmu$m along the anode wires and a
detection efficiency higher than 90\% have been achieved.
\end{abstract}


\begin{pacs}
29.40.Cs, 29.40.Gx \qquad {\bf DOI:} 10.1088/1674-1137/37/6/066002
\end{pacs}

\footnotetext[0]{\hspace*{-3mm}\raisebox{0.3ex}{$\scriptstyle\copyright$}2013
Chinese Physical Society and the Institute of High Energy Physics
of the Chinese Academy of Sciences and the Institute
of Modern Physics of the Chinese Academy of Sciences and IOP Publishing Ltd}%

\begin{multicols}{2}

\section{Introduction}

The production of {the} radioactive ion beam (RIB), which started in the
middle of {the} 1980s \cite{Tani85}, {opened} a new domain {in} nuclear
physics and astrophysics. Tens of radioactive ion beam lines have
been built worldwide. Two of them, the First Radioactive Ion Beam
Line in Lanzhou (RIBLL1) \cite{ribll1} and the Second Radioactive
Ion Beam Line in Lanzhou (RIBLL2) \cite{ribll2}, were constructed
at the Heavy Ion Research Facility in Lanzhou
(HIRFL) \cite{hirfl}, which is located at the Institute of Modern
Physics (IMP), the Chinese Academy of Sciences. RIBLL2 is a double
achromatic anti-asymmetry spectrometer to produce the RIBs with a
primary ion beam up to 1~GeV/u for either external target
experiment or storage ring mass spectroscopy. Normally, the type
of diagnostic detector used for tens of MeV/u RIBs is selected to
be a Parallel Plate Avalanche Counter (PPAC) \cite{ppac}  or a
low-pressure Multi-Wire Proportional {Chamber}
(MWPC) \cite{lpmwpc}. As for {RIBs} with energy of hundreds of
MeV/u or higher, these detectors will suffer {from} low
detection efficiency. In order to increase the detection
efficiency, a normal pressure MWPC is constructed for beam
diagnosis at RIBLL2. In the case of high vacuum of 10$^{-9}$ mbar
at RIBLL2, a special designed vacuum chamber to contain the MWPC
is constructed as well.

In this paper, the design of the MWPC and its readout will be
described, and the performance tested with 5.9~keV X-rays from a
$^{55}$Fe source and a 200~MeV/u $^{12}$C beam from CSRm will be also presented.

\section{Design and construction}
\subsection{Detector container}

Figure~\ref{fig1} shows the schematic layout of RIBLL2. It
consists of 4 dipoles and 20\, quadruple\, magnets,\, and\, its

\end{multicols}
\begin{center}
\includegraphics{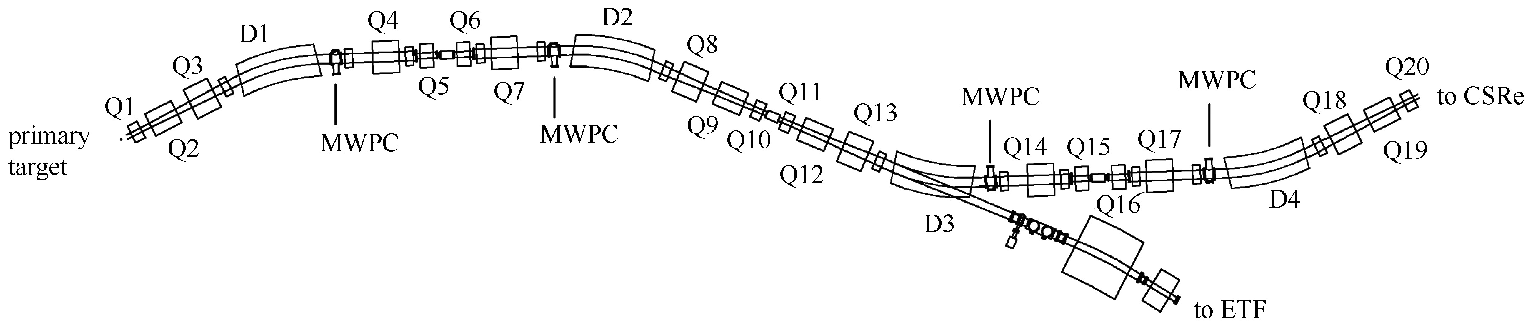}
\figcaption{\label{fig1} Schematic layout of RIBLL2 and mounting
positions of the MWPCs. }
\end{center}
\begin{multicols}{2}

\noindent total length is about 55~m. The primary beam extracted from
CSRm \cite{csr} bombards a target located at the entrance of
RIBLL2 to produce the radioactive ion beams through the projectile
fragmentation. The RIBs are purified and transported either to the
external target facility (ETF) or to CSRe for
experiments \cite{tu11}. In order to monitor the beam profile when
tuning the RIBLL2, MWPCs are employed. The positions where the
MWPCs are mounted are indicated in the figure.

In order to operate a normal-pressure MWPC in such an ultra high
vacuum of $10^{-9}$ mbar, each MWPC is mounted inside a movable
stainless steel pocket which is controlled by a step motor.  Fig.
2 shows {a} schematic drawing of the pocket. In order to sustain
one atmospheric pressure difference and to affect the beam as
slightly as possible, {0.1~mm} stainless steel foils are chosen
{for} the pocket {windows}. With such a device, the MWPC can be
moved in and out of the beam pipe freely.

\begin{center}
\includegraphics{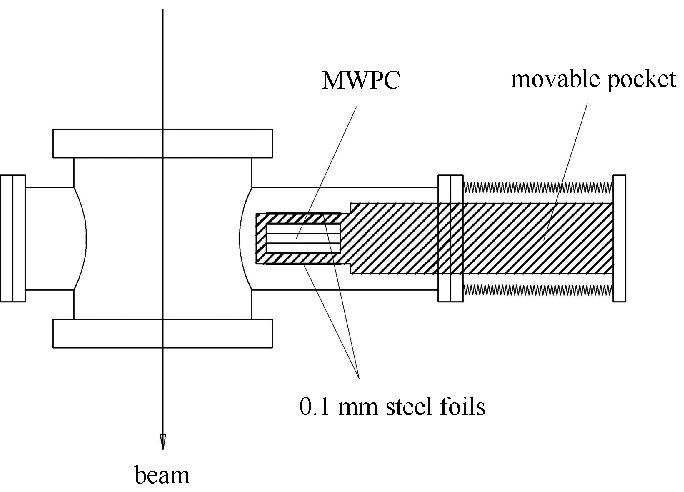}
\figcaption{\label{fig2} The special design to insert the
normal-pressure MWPC into the beam line with an ultra high vacuum.}
\end{center}

\subsection{Detector description}

Figure~\ref{fig3} shows the layout of the MWPC wire planes. The
anode plane is sandwiched between two cathode planes, and the
distance between the anode and each cathode is 3~mm. All of the
electrode frames are made from printed circuit board (PCB). The
active area of the whole MWPC is 80~mm$\times$80~mm. The anode
plane consists of 20 Au-W parallel arranged wires of 25~$\upmu$m in
diameter with 4~mm spacing. In order to avoid the dramatically
increasing electric field and reduce the spark at the boundaries
of the chamber, the traditional way is to substitute the end wires
with protection wires in larger diameters. However, this could
shrink the sensitive area of the detector. During the research, we
found the gain at the boundaries would not increase if the end
wires were away from the frame. Therefore, the inner length of the
frames is designed as 84~mm$\times$84~mm, and the distance
between the end wires and the inner edge of the frame is 2~mm. All
the anode wires are soldered together and only one fast time
signal output is provided to trigger the data acquisition (DAQ)
system. Each of the cathode planes consists of 80 Be-Cu parallel
arranged wires of 75~$\upmu$m in diameter with 1~mm spacing. Every
two adjacent cathode wires are soldered together to form one
readout strip of 2~mm in width. Both cathodes are placed
orthogonally to each other. All the wires are positioned and
tensed with a programmable winding machine. The accuracy of the
positioning of the wires is about 25~$\upmu$m, and the tension
provided on all wires is 70~g.

For extracting the signals from the cathodes, the delay line
technique is employed. The commercial DIP delay line module 1520
from Data Delay Devices company \cite{ddd} is selected. Each
module has {10 taps} fixed {onto it}, and 4 modules are used for one cathode
plane in our case.

\begin{center}
\includegraphics{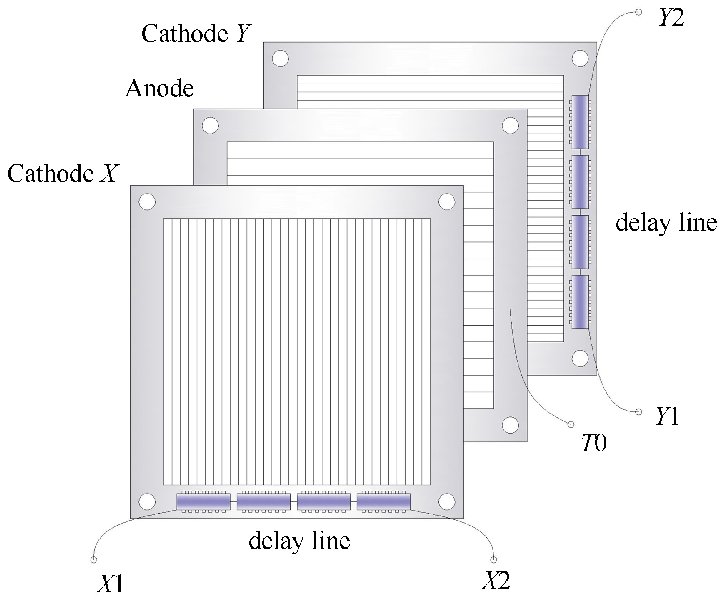}
\figcaption{\label{fig3} Structure of the MWPC used for beam diagnostics. }
\end{center}

From the viewpoint of spatial resolution, it is better to choose
a delay line with longer delay time, but this will decrease the
efficiency. We chose 40~ns delay line modules to balance the
spatial resolution and the efficiency. Table~\ref{tab1} lists the
characteristics of the module. The nominal delay time between two
adjacent readout wires is 4~ns, which corresponds to a 2~mm
distance, and the nominal tolerance is 0.2~ns. The intrinsic
position uncertainty caused by the delay line is estimated to be 50~$\upmu$m.

\begin{center}
\tabcaption{ \label{tab1}  Characteristics of the delay line module.}
\footnotesize
\begin{tabular*}{86mm}{@{\extracolsep{\fill}}cccc}
\toprule
& parameter & value\\
\hline
& number of taps & 10 \\
& total delay & (40$\pm$2)~ns\\
& delay between taps & (4$\pm$0.2)~ns\\
& characteristic impedance & 50~$\Omega$ \\
& 3dB bandwidth & 43.75~MHz \\
& rising time & 8~ns \\
\bottomrule
\end{tabular*}
\vspace{0mm}
\end{center}

With the delay line readout technique, only two output signals are
needed for each cathode (see also Fig.~\ref{fig3}). The position
can be calculated with the following equations:
\begin{align}
x&=k_x(T_{x1}-T_{x2}),\\[1mm]
y&=k_y(T_{y1}-T_{y2}),
\end{align}
where $T_{x1}$ and $T_{x2}$ are the time from both ends of Cathode
$X$, $T_{y1}$ and $T_{y2}$ are the time of Cathode $Y$, and $k_x$ and
$k_y$ are constant values depending on the width of detector and
the electronics, respectively.

\section{Test results}

The performance of the assembled MWPCs has been tested with
{a} radioactive ion source and the beam. The position (spatial)
resolutions of the MWPCs were tested with a 5.9~keV X-ray
$^{55}$Fe source, and their detection efficiencies were tested
with the slow extracted $^{12}$C beam of 200~MeV/u from CSRm.

\subsection{\boldmath Test with a $^{55}$Fe source}

During the test, the signals from the anode and two cathodes are
first sent into a fast preamplifier FTA810, and then a constant
fraction discriminator CF8000. After that, the cathode signals are
input into a PXI TDC module \cite{Liu09} based on the
HPTDC \cite{hptdc} chips to record the time information. The DAQ
system is triggered by the anode signal.

The MWPC was operated at a high voltage of 1950~V under normal
pressure with the gas flow mode. The filling gas was a mixture of
80\% Ar and 20\% CO$_2$, and the gas flow rate was about 1.8~l/h.
During the test, a collimator with 3 slits was used for position
calibration. Each slit is 0.18~mm wide and 1.8~mm apart from each
other. The $^{55}$Fe source and the collimator moved along the
anode wires step by step under the control of a step motor with a
step length of 5.4~mm and an accuracy of 25~$\upmu$m.

Figure~4(a) shows the measured position spectrum, in which each peak
represents the position of one of the slits. By fitting the
positions the performance of the MWPC detector can be evaluated.
Fig.~4(b) shows both the obtained integral nonlinearity (INL) and
differential nonlinearity (DNL). In the full range of this
detector, the INL is within 0.42~mm (0.5\%) and the DNL is within
0.33~mm (0.4\%). Fig.~4(c) shows the corresponding position
resolution distribution. Obviously, the spatial resolution is
better than 250~$\upmu$m in the full detector range, and the average
value is measured to be $\sigma$=(186$\pm$32)~$\upmu$m.

\subsection{\boldmath Test with a 200 MeV/u $^{12}$C beam}

To obtain the detection efficiencies of the MWPC detectors, three
MWPCs were mounted sequentially in one of the beam lines of the
CSRm. The distance between each two MWPCs was 15 cm. A $^{12}$C
beam with an energy of 200 MeV/u was used for the test. The
electronics used {were} similar to {those} in the X-ray test, except
that a downstream plastic scintillator signal was used to trigger the DAQ system.

\begin{center}
\includegraphics{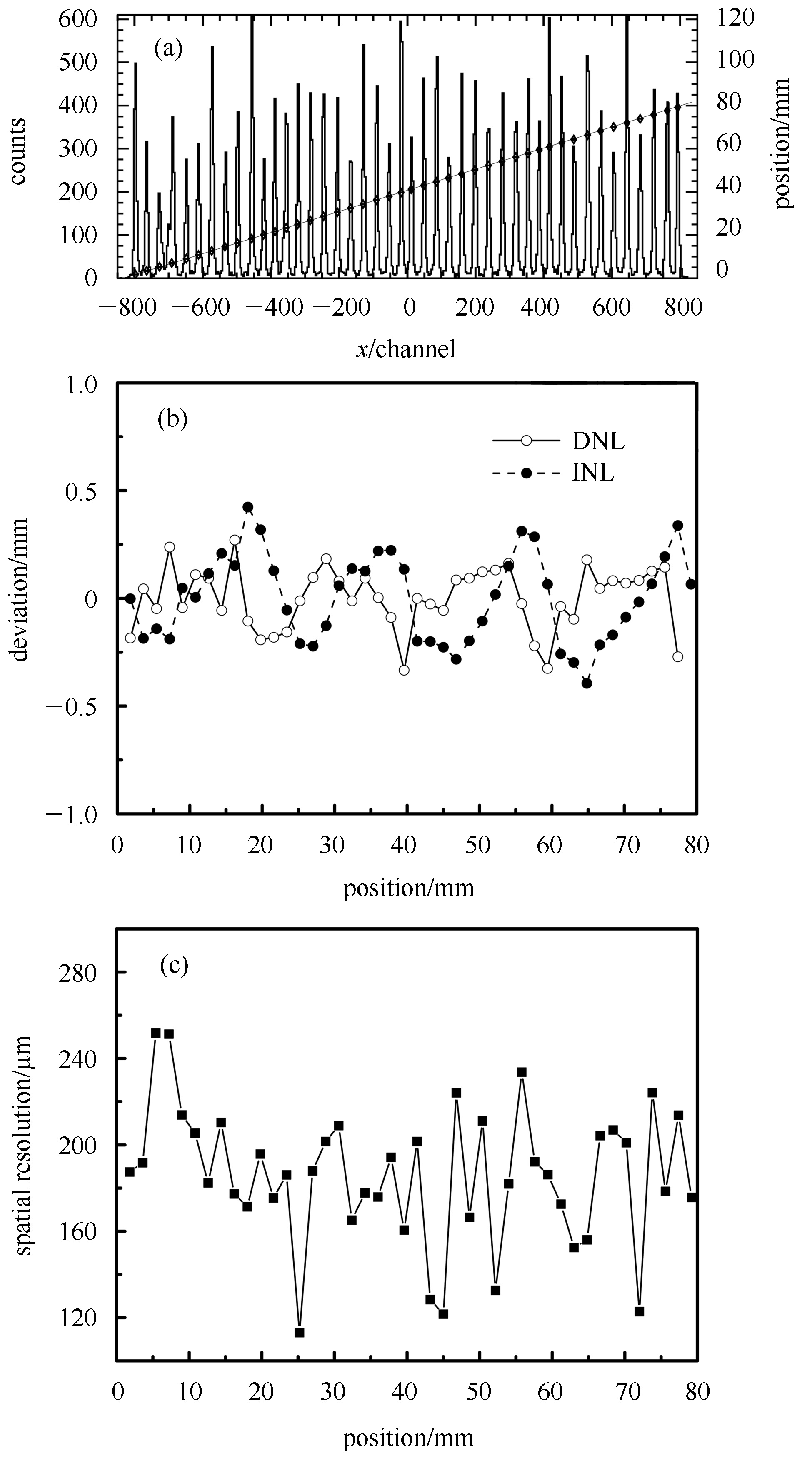}
\figcaption{\label{fig5} Position spectrum (a), nonlinearities (b)
and spatial resolution (c) measured with a $^{55}$Fe source. INL
represents the integral nonlinearity, and DNL the differential
nonlinearity, respectively.}
\end{center}

Figure~5(a) shows a typical $^{12}$C beam profile monitored with a
MWPC. The lines parallel to the $x$ axis represent the anode wires.
The position resolution was determined in the following way. For $x$
direction, we can calculate\, the\, position of a\, charged\, particle\, in
MWPC$_2$

\end{multicols}
\begin{center}
\includegraphics{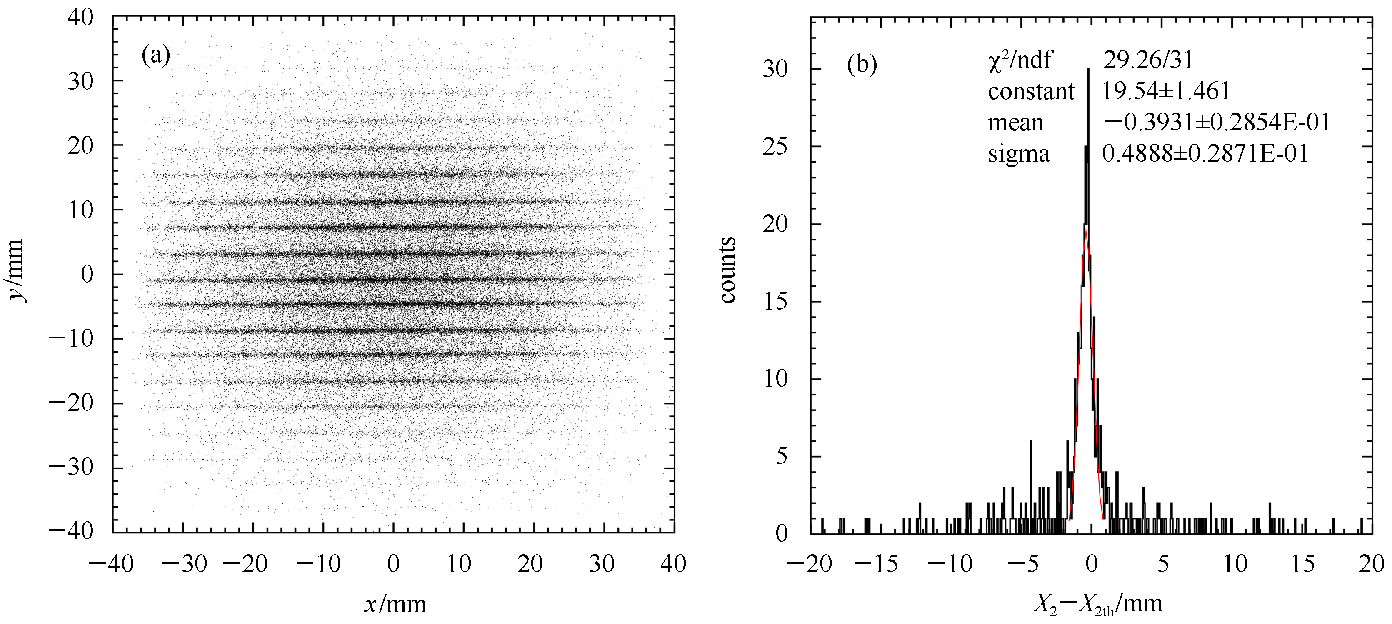}
\figcaption{\label{fig6} (a)Beam profile of $^{12}$C monitored by
the MWPC. (b)Position resolution on $X$ plane.}
\end{center}
\begin{multicols}{2}

\noindent with the trajectory determined by MWPC$_1$ and MWPC$_3$,
which is labeled as $x_{\rm 2th}$, then the difference between
$x_{\rm 2th}$ and $x_{2}$ is plotted in Fig.~5(b). The resolution is
$\sigma_{x0}$=(489$\pm$29)~$\upmu$m, which is contributed by all
the three MWPCs. Therefore, the position resolution for one MWPC
is $\sigma_x$=(282$\pm$17)~$\upmu$m. As for $y$ direction, the
position resolution is $\sigma_y$=$s/\sqrt{12}$=1.15~mm.

The detection efficiency can be calculated by the following equations:
\begin{align}
\eta_x&=\frac{N_{123x}}{N_{13x}},\\[3mm]
\eta_y&=\frac{N_{123y}}{N_{13y}},
\end{align}
where $N_{13}$ is the number of events detected with the $1^{\rm st}$
and $3^{\rm rd}$ MWPC (to assure the ion has passed all the three
MWPCs), $N_{123}$ is the number of events detected by the $2^{\rm nd}$
MPWC among all the $N_{13}$ events, and the indices $x$ and $y$
represent $x$ and $y$ directions, respectively.

We applied 1855~V bias for the detectors and tested them with the
beam intensity as high as 10$^5$~pps. The detection efficiency we
achieved was 94.9\% for $x$ direction and 90.1\% for $y$ direction.
This performance fully reaches the requirement of the beam
diagnostics at RIBLL2.

\section{Summary}

We have developed a normal-pressure MWPC for beam diagnostics at
RIBLL2. The MWPC consists of three-layer wire planes and can work
at normal pressure with a special designed container. We have
tested the MWPC with a $^{55}$Fe X-ray source and achieved a very
good position resolution and linearity. The spatial resolution
along the anode wires for the whole detector is better than
250~$\upmu$m. The integral nonlinearity is within 0.5\% and the
differential nonlinearity is within 0.4\%. We have also tested the
MWPC with a 200~MeV/u $^{12}$C beam at the intensity of $10^5$~pps
and get the detection efficiency higher than 90\%. The tests show
that the MWPC can work properly as a beam diagnostic detector for
$Z\geqslant$6 beams with energy $\leqslant$200~MeV/u. In the future, we will
test it with lighter and higher-energy beams.

\end{multicols}

\vspace{-2mm}
\centerline{\rule{80mm}{0.1pt}}
\vspace{2mm}

\begin{multicols}{2}


\end{multicols}

\clearpage

\end{document}